**Structural and Magnetic Properties of Single Crystals of the Geometrically Frustrated Zirconium Pyrochlore, $Pr_2Zr_2O_7$**

M. Ciomaga Hatnean[1*], C. Decorse[2], M. R. Lees[1], O. A. Petrenko[1], D. S. Keeble[1] and G. Balakrishnan[1]

[1] *Physics Department, University of Warwick, Coventry, CV4 7AL, UK*
[2] *SP2M-ICMMO, UMR 8182, Universite Paris-Sud 11, F-91405 Orsay, France*

Abstract

We report the growth of large high quality crystals of the frustrated magnet, $Pr_2Zr_2O_7$ by the floating zone technique. Several crystals have been produced starting with varying levels of excess of $Pr_6O_{11}$ to compensate for the loss of $Pr_2O_3$ during the growth process. X-ray diffraction analysis of the crystals shows that the Pr composition in our crystals, as determined by the Pr site occupancy in the pyrochlore structure, is always slightly less than 1. Fits to the magnetic susceptibility data below 10 K using a Curie-Weiss law reveal an antiferromagnetic coupling between the Pr moments, with an estimated Weiss temperature of $\Theta_W$ = -2.5 K and an effective moment for the $Pr^{3+}$ of $\mu_{eff}$ = 2.5 $\mu_B$. Magnetization versus field measurements show a spin-ice like anisotropy.

1.  Introduction

Geometrical frustration arising from the arrangement of the spins in the lattice is an important feature in magnetism. One model structure used to study 3D geometrical magnetic frustration is the pyrochlore, which consists of corner sharing tetrahedra [1]. Geometrically frustrated pyrochlore oxides of the type $A_2B_2O_7$ (where A=rare earth, B=Ti or Zr) have been the subject of extensive investigations because of their interesting and rather unconventional magnetic properties, such as spin liquid [2], spin glass [3, 4], or spin ice behaviour [5, 6]. Frustrated magnetism in the oxide pyrochlores has been widely investigated using techniques such as magnetic susceptibility, specific heat and neutron diffraction down to low temperatures [7, 8]. Although the magnetic properties and ground states of the titanate pyrochlores have been investigated in detail, similar studies of the zirconate pyrochlores have been hampered by the lack of large, good quality single crystals. With the exception of a few investigations on $Pr_2Zr_2O_7$ single crystals [9-11], most of the research on the zirconate pyrochlores has used powder samples [12-19].

High quality single crystals are necessary to investigate the interesting properties of these zirconate systems. Single crystals of zirconates can be grown by the floating zone technique, a technique that has already been successfully applied to the growth of the titanate pyrochlores [20]. The melting points of the zirconate pyrochlores are higher than those of the titanates and it is therefore necessary to use high power lamps when using the optical floating zone technique for the crystal growth. Previous studies of the crystal growth of $Pr_2Zr_2O_7$ [9] have shown that one consequence of this is the evaporation of $Pr_2O_3$ during the crystal growth process and that this can cause a reduction in the Pr content in the single crystals. Controlling any changes in stoichiometry and the subsequent determination of the cationic content is therefore crucial for the study and the understanding of the magnetic properties of these materials, due to their direct dependence on the A/B (Pr/Zr) site occupancy/ratio [21].

In this paper, we report a systematic study of the crystal growth of $Pr_2Zr_2O_7$ by the floating zone technique using a high power xenon arc lamp furnace. Structural characterization and investigation of the magnetic properties of this new class of geometrically frustrated zirconium pyrochlore oxides have also been carried out. Several high-quality single-crystals of $Pr_2Zr_2O_7$ have been prepared, starting with different amounts of excess $Pr_6O_{11}$ to compensate for any losses of Pr during crystal growth. Powder and single-crystal X-ray diffraction have been used to investigate the structural properties of the crystals produced and compositional analysis has also been carried out to estimate the Pr content in the as grown crystals. In order to study the magnetic behaviour of our crystals, we have carried out *dc* magnetic susceptibility measurements as a function of temperature (1.8 ≤ *T* ≤ 300 K) and magnetic field (*T* = 1.45 K, *H* ≤ 110 kOe). These measurements were performed with the field applied along three different high symmetry directions.

2.  Experimental Methods

Polycrystalline samples of $Pr_2Zr_2O_7$ were prepared by the standard solid state reaction. To compensate for the evaporation of $Pr_2O_3$ during the growth previously reported by K. Matsuhira *et al.* [9], single crystals of $Pr_2Zr_2O_7$ were grown using different starting compositions of the feed rods: a stoichiometric composition and two compositions rich in $Pr_6O_{11}$ (5 and 10% $Pr_6O_{11}$ excess). Stoichiometric quantities of 99.9% purity $Pr_6O_{11}$ and $ZrO_2$ powders were mixed, ground, and heated to 1400 °C in air for several days with intermediate grindings. The rods required for crystal growth (feed and seed rods) were prepared from the reacted stoichiometric powder obtained, to which the required amounts of excess $Pr_6O_{11}$ (0, 5 and 10%) were added prior to the final sintering. Cylindrical rods, 6-8 mm in diameter and about 70-80 mm long, were produced by isostatic compression and then sintered at 1500 °C in air in preparation for the crystal growth. Powder X-ray diffraction measurements were used to check the composition and structure of these rods prior to crystal growth. The crystals were grown by the floating zone technique [9] using a four-mirror Xenon arc lamp optical image furnace (CSI FZ-T-12000-X_VI-VP, Crystal Systems Incorporated, Japan). The crystal growths were performed in oxygen at pressures in the range 1-4 bars. All the crystals were prepared using a growth rate of 15 mm/h. The feed and the seed rods were counter-rotated at 20-30 rpm to ensure efficient mixing and homogeneity. Initially, polycrystalline rods were used as seeds and once good quality crystals were obtained, a crystal seed was used for subsequent growths.

---

[*] Corresponding author
    *Email address:* M.Ciomaga-Hatnean@warwick.ac.uk

Single phase identification of our crystals and analysis of the Pr composition was performed at room temperature by powder X-ray diffraction (XRD) on a Panalytical X-ray diffractometer using Cu K$\alpha_1$ radiation ($\lambda$ = 1.5406 Å). A small quantity of each boule was ground into powder and the data were then collected between 10 and 110° 2θ, with a step size of 0.013° in 2θ, and a total scanning time of 12 hours. The X-ray data were analysed with the FullProf software suite [22]. To confirm the crystal structure of $Pr_2Zr_2O_7$, single crystal X-ray diffraction measurements were performed on the grown crystals. Small blocks of the crystals prepared using different starting compositions (0%, 5% and 10% $Pr_6O_{11}$ excess) (with dimensions of ~100 μm × ~100 μm × ~60 μm) were cleaved from the bulk and mounted on a Gemini R CCD X-ray diffractometer (Agilent Technologies). Data over large regions of reciprocal space were collected using MoK$\alpha$ radiation ($\lambda$ = 0.7107 Å) at room temperature. The data were indexed and integrated using CrysAlisPro (Agilent Technologies). The structure was refined using ShelX [23]. Composition analysis on the crystal grown from the rod made from the stoichiometric starting material was carried out by energy dispersive X-ray spectroscopy (EDAX) using a scanning electron microscope. Several pieces were cut from the boule along the growth direction (beginning, middle and end) in order to investigate the chemical composition along the entire length of the crystal boule.

The quality of the as-grown single crystals was checked and the samples were aligned for cutting using a real time Laue X-ray imaging system with a Photonic-Science Laue camera system. Rectangular prism shape samples with [100], [110], and [111] directions perpendicular to the largest face were cut out for the magnetization measurements from the boule grown with the stoichiometric starting composition.

In order to study the effects of annealing in a reducing atmosphere on $Pr_2Zr_2O_7$, a few pieces of the aligned single crystals taken from the sample prepared using a starting composition with 0% $Pr_6O_{11}$ excess were annealed in Ar (3% $H_2$) flow at 1200°C for 2 days.

Preliminary neutron scattering experiments were carried out at LLB-Orphée (CEA-Saclay, France) on the thermal neutron four-circle spectrometer 6T2 and on the cold neutron triple-axis spectrometer 4F.

Magnetization measurements were made using a Quantum Design MPMS-5S SQUID magnetometer and an Oxford Instruments vibrating sample magnetometer (VSM). The magnetization was measured as a function of temperature in a constant magnetic field of 1 kOe from 1.8 to 350 K. The *M(H)* curves at a fixed temperature of 1.45 K were also measured in magnetic fields of up to 110 kOe. The field was swept at a rate of 2.5 kOe/min and the data were collected every 40 Oe.

3. Results and Discussion

The powder X-ray diffraction pattern obtained for the feed rod of the stoichiometric composition is shown in Figure 1. An analysis of the pattern made using the FullProf software suite provided a good fit to the pyrochlore structure (space group $Fd\bar{3}m$), with a lattice parameter of 10.70988(3) Å. This value is in agreement with the previously published value of 10.72 Å for $Pr_2Zr_2O_7$ [9]. There are no impurity peaks present in the pattern.

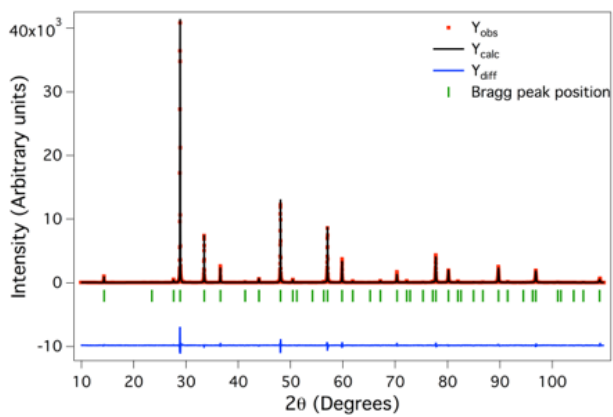

Figure 1. Room temperature powder X-ray diffraction pattern (in red) of the stoichiometric polycrystalline rod of $Pr_2Zr_2O_7$ used for the crystal growth. Also shown are the fit to the cubic $Fd\bar{3}m$ structure (in black) and the difference curve (in blue).

As expected, the X-ray diffraction patterns collected on powder from the sintered rods used for the growth of the crystals containing 5% and 10% $Pr_6O_{11}$ excess showed these materials are a mixture of phases (praseodymium oxide and praseodymium zirconate).

The crystal growths carried out using the different starting compositions (stoichiometric, 5% excess $Pr_6O_{11}$, and 10% excess $Pr_6O_{11}$) all produced crystals of good quality. The crystal growth of the stoichiometric composition was performed in oxygen at a pressure of 1 bar, while for the growths of the crystals using the $Pr_6O_{11}$ rich compositions, the pressure used was increased to between 3.5 and 4 bars in order to limit the loss of the excess $Pr_6O_{11}$. Despite the use of higher gas pressures, we observed an increase in the evaporation of $Pr_2O_3$ for the crystals grown with a $Pr_6O_{11}$ excess (both 5 and 10%), as the amount of the brown-coloured deposition observed on the quartz tube at the end of those crystal growths was larger.

Most of the crystals developed facets as they grew and all of them had a dark brown colour. Figure 2 shows photographs of the as grown crystals of $Pr_2Zr_2O_7$ prepared by the floating zone method, using a starting composition of 0% and 10% $Pr_6O_{11}$ excess. The crystals were typically 5 to 8 mm in diameter and 40 to 85 mm long. The quality of the crystals was investigated by X-ray Laue diffraction. The Laue patterns indicate that the crystals had no preferred growth direction. An X-ray Laue photograph taken of a crystal of $Pr_2Zr_2O_7$ prepared using a starting composition of 0% $Pr_6O_{11}$ excess is shown in Figure 3 and confirms the high quality of the crystals grown. Crystals annealed in Ar (3% $H_2$) flow become bright green coloured, as shown in the inset of Figure 2a. It has been suggested previously [24], that the change in colour is associated with a change of the oxidation state of the Pr ion, from $Pr^{4+}$ to $Pr^{3+}$.

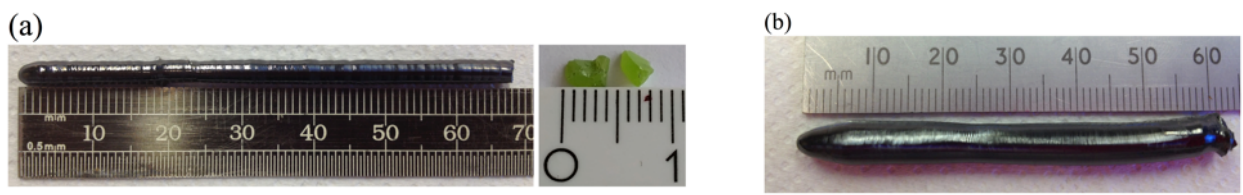

Figure 2. Single crystals of $Pr_2Zr_2O_7$ grown by the floating zone method, (a) part of a boule grown from a stoichiometric starting composition (with no $Pr_6O_{11}$ excess), in 1 bar oxygen pressure at 15 mm/h. Inset: a few pieces from the boule annealed in Ar +3% $H_2$ flow for 2 days, (b) boule grown from a starting composition with a 10% $Pr_6O_{11}$ excess in 4 bar oxygen pressure at 15 mm/h.

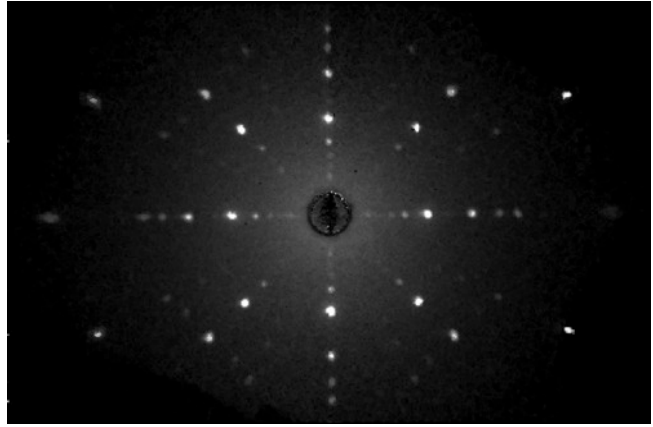

Figure 3. X-ray Laue back reflection photograph of a crystal of $Pr_2Zr_2O_7$, grown from a starting composition with no $Pr_6O_{11}$ excess, showing the [100] orientation.

Composition analysis by EDAX of the $Pr_2Zr_2O_7$ crystal produced from the rod made with the stoichiometric starting material shows that the cationic ratio averages to 2:2 for Pr:Zr over the entire length of the crystal boule, despite the observed evaporation during the crystal growth. The average atomic percentages of Pr and Zr were (18.7 ± 0.1) % and (17.6 ± 0.1) % respectively, and were found to be very close to the theoretical value of 18.2 % expected for both the Pr and the Zr.

To further investigate the quality of the crystals and their suitability for future neutron scattering experiments, the crystal prepared using the stoichiometric starting composition was examined using single-crystal neutron diffraction. Data collected on the cold neutron triple-axis spectrometer 4F installed at the LLB-Orphée with the crystal aligned along the [111] direction reveal that our $Pr_2Zr_2O_7$ crystal has a mosaicity of ~ 0.75°.

Room temperature X-ray diffraction data were also collected on small portions of the crystals that were powdered. The X-ray data were analysed using the Rietveld method in order to obtain detailed information about the crystallographic structure of our $Pr_2Zr_2O_7$ crystals.
The results of the Rietveld refinements obtained for the X-ray data on the $Pr_2Zr_2O_7$ ground crystals, show no impurity phases are present in any of our $Pr_2Zr_2O_7$ crystals. The lattice parameters increase gradually with increasing $Pr_6O_{11}$ excess used to prepare the feed rods. The crystal grown with a stoichiometric composition shows the lowest value of the lattice parameter ($a$ = 10.67795(9) Å), while for the crystals prepared using starting compositions enriched in $Pr_6O_{11}$ the calculated values were of 10.70822(7) Å (5% excess) and 10.72081(7) Å respectively (10% excess). The lattice parameters of the samples which were prepared using starting compositions with 5 and 10% $Pr_6O_{11}$ are very similar to the previously reported values, although the starting compositions of the rods used to grow these crystals are not given in the paper [10].

The $A_2B_2O_7$ compounds, where $A^{3+}$ is a rare earth cation and $B^{4+}$ is Ti or Zr cation, crystallise in a cubic structure with the $F\bar{d}3m$ space group and there are eight molecules per unit cell ($Z$ = 8). The structure is composed of two types of cation coordination polyhedron. The bigger A cations are eight coordinated and are situated on the *16d* sites, within scalenohedra (distorted cubes) that contain six equally spaced anions (O2 atoms) at a slightly shorter distance from the central cations [4]. The O2 anions occupy the *48f* sites, the only variable positional parameter of the structure. The smaller B cations are six coordinated and are situated on the *16c* sites, located in the centre of O1 oxygen trigonal antiprisms, with all the six anions at equal distances from the central cations [1]. The O1 atoms occupy the *8b* sites.

In order to obtain the atomic positions, the Debye-Waller factors and the values of the site occupancies for the Pr and Zr sites, small pieces of the as grown $Pr_2Zr_2O_7$ single-crystals prepared from the different starting composition (0% , 5% and 10% $Pr_6O_{11}$ excess) were cleaved from the boules and room temperature single crystal diffraction data were collected. For the $Pr_2Zr_2O_7$ single-crystal prepared from a starting composition with 0% $Pr_6O_{11}$ excess, 8286 reflections were measured, of which 134 were independent reflections. The structure was then refined using the ShelX software. An extinction parameter was refined and the obtained value was 0.0026(3). Attempts to fit the single-crystal X-ray data using different models for the occupancies the Pr and the Zr sites, showed that the best model for the refinements is one in which only the Pr occupancy varies. The results of the refinement are summarised in Table 1 and the crystallographic structure of $Pr_2Zr_2O_7$ is shown in Figure 4.

|     | x | y | z | $U_{iso}*/U_{eq}$ | Occ. |
| --- | --- | --- | --- | --- | --- |
| Pr | 0.5000 | 0.5000 | 0.5000 | 0.0062(4) | 0.974(9) |
| Zr | 0 | 0 | 0 | 0.0065(5) | 1 |
| O1 | 0.3750 | 0.3750 | 0.3750 | 0.0098(16) | 1 |
| O2 | 0.3354(5) | 0.1250 | 0.1250 | 0.0144(11) | 1 |
|     | $U_{11}$ | $U_{22}$ | $U_{33}$ | $U_{12}$ | $U_{13}$ | $U_{23}$ |
| Pr | 0.0062(4) | 0.0062(4) | 0.0062(4) | -0.00151(7) | -0.00151(7) | -0.00151(7) |
| Zr | 0.0065(5) | 0.0065(5) | 0.0065(5) | 0.00157(3) | 0.00157(3) | 0.00157(3) |
| O1 | 0.0098(16 | 0.0098(16) | 0.0098(16) | 0 | 0 | 0 |
| O2 | 0.019(2) | 0.0121(13) | 0.0121(13) | 0 | 0 | 0.0031(15) |

Table 1. Refined structural parameters obtained for single-crystal $Pr_2Zr_2O_7$ prepared from powder with 0% excess $Pr_6O_{11}$. Atomic positions for the $Fd\bar{3}m$ (origin choice 2) cubic structure are for Pr, 16d (0.5, 0.5, 0.5); Zr, 16c (0, 0, 0); O1, 8b (0.375, 0.375, 0.375); and O2, 48f (x, 0.125, 0.125). The goodness of the fit is of 1.367, with $R_w$= 0.0581 and $R$= 0.0216.

The Pr site occupancy and the *x* position of the O2 atoms single-crystal X-ray diffraction data are in agreement with those of the previous work of K. Matsuhira *et al.* (Pr site occupancy equal to 1.00) [9] and K. Kimura *et al.* (Pr site occupancies equal to 0.998(3) and 0.968(3) for two different crystals) [10].

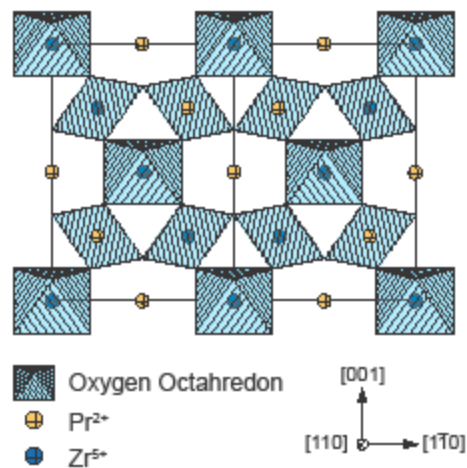

Figure 4. Crystallographic structure of $Pr_2Zr_2O_7$, as obtained from the single crystal X-ray diffraction data refinement.

The results of the Rietveld refinements (the atomic position for O2 and the Debye-Waller factors for all atoms) obtained for the single-crystal X-ray data on the $Pr_2Zr_2O_7$ crystals prepared with starting composition using 5% and 10% $Pr_6O_{11}$ excess were similar to those of the stoichiometric composition crystal. The Pr site occupancy obtained for the single-crystals prepared with $Pr_6O_{11}$ enriched starting compositions are found to be 0.973(8) (for 5% excess) and 0.975(7) (for 10% excess). The goodness of the fit was of 1.461, in both cases.
The refinements of the single crystal X-ray diffraction data confirm that all the crystals grown show a very small deficiency in Pr content regardless of the $Pr_6O_{11}$ excess used for the crystal growth process. Given the very similar Pr:Zr ratios in all of the crystals grown, we have restricted our measurements of the magnetic properties along the three different high symmetry directions, described below, to the crystal grown from the stoichiometric starting composition.

We also tried the crystal growth of $Pr_2Zr_2O_7$ starting with feed rods made with $Pr_6O_{11}$ enriched powders where the $Pr_6O_{11}$ excess is added at the start of the preparation process, before the first solid state reaction of the powder, i.e., mixtures with the starting compositions $Pr_{2.2}Zr_2O_7$ (5% excess) and $Pr_{2.4}Zr_2O_7$ (10% excess) were reacted. The feed rods obtained were then used for crystal growth. In these cases, the growths were unstable and the resulting boules were of inferior quality.

Field-cooled (FC) and zero-field-cooled (ZFC) magnetization versus temperature curves were measured on oriented $Pr_2Zr_2O_7$ crystals prepared using a stoichiometric starting composition. Figure 5 shows the temperature dependence of the *dc* magnetic susceptibility, $\chi(T)$ and reciprocal *dc* magnetic susceptibility $\chi^{-1}(T)$ measured along three different crystallographic directions. $\chi(T)$ data measured along all three high symmetry directions exhibits a monotonic increase upon cooling from 350 to 1.8 K. No anomaly has been observed up to 1.8 K, indicating the absence of a magnetic transition. Overall, the magnetic susceptibilities collected along the [110] and [111] directions appear to overlap in the entire temperature range of 350 to 1.8 K, while the magnetic susceptibility along the [100] direction is clearly higher in magnitude (see Figure 5a). However, at lower temperatures, $\chi^{-1}(T)$ data reveal that the susceptibilities measured along the three high symmetry directions are clearly different in terms of magnetic behaviour (see Figure 5b). A strong local Ising magnetic anisotropy is confirmed by the local susceptibility measurements made using polarised neutron scattering at LLB-Orphée [25]. Our magnetic susceptibility results are slightly different than those published by K. Kimura *et al.* [11], whose $\chi(T)$ show an isotropic magnetic response at higher temperature, although their neutron scattering experiments also show a <111> Ising anisotropy for $Pr_2Zr_2O_7$ [10]. The difference in the magnetic behaviour observed by different techniques, analogous to that found in $Tb_2Ti_2O_7$, comes from the fact that magnetic susceptibility measurements, even on single crystals average over the four local anisotropy axes of a Pr tetrahedron [26].

Attempts to fit to the $\chi^{-1}$ vs. *T* data show that the $\chi^{-1}(T)$ data do not obey a Curie-Weiss law in the temperature range 1.8 to 350 K. Nevertheless, we have made fits to Curie-Weiss law to compare with the analysis used in the previous work [9, 10] and the results of these fits are summarised in Table 2. The $\chi^{-1}$ vs. *T* fitted data in the temperature range of 1.8 to 10 K point to an antiferromagnetic coupling between the Pr spins, although we obtain a slightly larger value of the estimated Weiss temperature, $\Theta_W$, compared to the previous reports [9, 10]. This difference may be explained by the fact that our measurements were performed on an orientated crystal, along well-defined crystallographic axes, while the previous results were for randomly oriented crystals [9]. In order to check this, we have repeated our measurements on a ground piece of our $Pr_2Zr_2O_7$ crystal and we have confirmed that the estimated Weiss temperature for the measurement on powder gives a slightly smaller value for $\Theta_W$, (see Table 2). The difference between the data collected on the ground crystal and along the three high symmetry directions cannot be explained by the contribution of the demagnetizing factors; when considering the demagnetizing factors, the difference in the calculated susceptibility, for all the crystallographic directions, has been found close to 1%. In addition, we calculate the effective moment for all our crystals and found that the values are in agreement with the recent published results on $Pr_2Zr_2O_7$ [10].

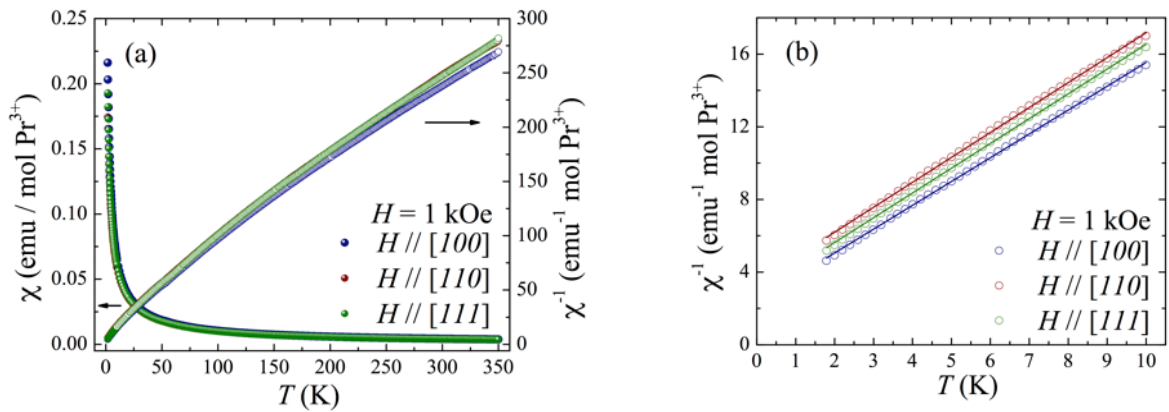

Figure 5. (a) Temperature dependence of *dc* magnetic susceptibility, $\chi$ vs. *T* and the reciprocal *dc* magnetic susceptibility, $\chi^{-1}$ vs. *T* of a crystal of $Pr_2Zr_2O_7$, with a magnetic field applied along the [100], [110], and [111] directions in the temperature range 1.8 to 350 K. (b) $\chi^{-1}$ vs. *T* and lines showing linear fits to the $\chi^{-1}$ vs. *T* data, along the [100], [110], and [111] directions in the temperature range 1.8 to 10 K.

| Pr$_6$O$_{11}$ excess (%) | Orientation | $\Theta_W$ (K) | $\mu_{eff}$ ($\mu_B$) |
|---|---|---|---|
| 0 | [100] | -1.85 ± 0.02 | 2.470 ± 0.004 |
| | [110] | -2.51 ± 0.02 | 2.414 ± 0.005 |
| | [111] | -2.12 ± 0.02 | 2.421 ± 0.005 |
| | powder | -1.72 ± 0.02 | 2.547 ± 0.006 |
| 5 | powder | -0.74 ± 0.01 | 2.562 ± 0.003 |
| 10 | powder | -0.70 ± 0.02 | 2.565 ± 0.003 |
| Previous studies | unknown orientation[9] [111][10] | -0.55[9] -1.4[10] | 2.5(1)[10] |

Table 2. Weiss temperatures and effective moment values obtained for the fit to the $\chi^{-1}$ vs. *T* data collected on the Pr$_2$Zr$_2$O$_7$ single-crystals grown using different starting compositions (0, 5 and 10 % Pr$_6$O$_{11}$ excess). The results obtained along different crystallographic directions on the Pr$_2$Zr$_2$O$_7$ single-crystal grown using a stoichiometric starting composition are also presented.

Magnetization as a function of applied magnetic field *M(H)* at 1.45 K was measured on the Pr$_2$Zr$_2$O$_7$ crystal prepared using a stoichiometric starting composition. Figure 6 shows the field dependence of the magnetization measured along the [100], [110], and [111] directions. The data reveals a non-linear variation of the magnetization as a function of applied field. These results are similar to those reported previously for Pr$_2$Zr$_2$O$_7$ [11], showing that the magnetization follows a spin ice like behaviour, with a different magnetic-field dependence along different crystallographic directions. However, the magnetization values are smaller than those previously reported, which may be explained by a different Pr/Zr ratio of our sample. This is supported by the different $\Theta_W$ and $\mu_{eff}$ values obtained by the fits performed on the magnetization data collected along the [111] direction (see Table 2). The anisotropy observed in the magnetization process is a key feature of a spin-ice system, as in the Dy$_2$Ti$_2$O$_7$ and Ho$_2$Ti$_2$O$_7$ pyrochlore oxides [27, 28]. The *M(H)* measurements indicate that the magnetic anisotropy is compatible with the spin-ice model only to a certain degree, as the magnetization is not fully saturated even in an applied magnetic field of 110 kOe, and only 77-83% of the expected saturated moments for a classic spin-ice configuration [27] is recovered.

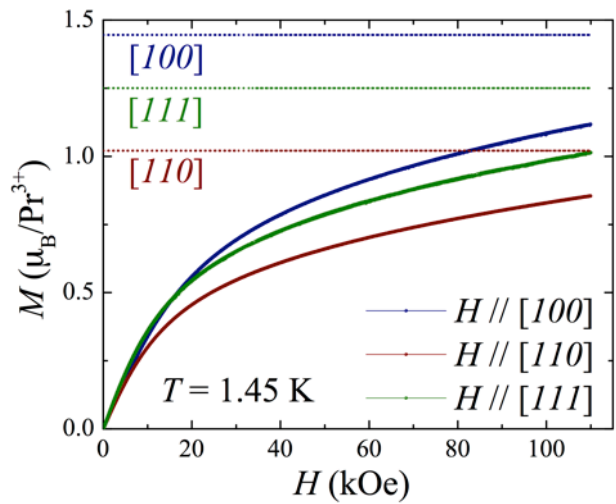

Figure 6. Isothermal magnetization (*M*) at *T* = 1.45 K of a crystal of Pr$_2$Zr$_2$O$_7$, along the [100], [110], and [111] directions, plotted as a function of applied magnetic field (*H*).

Magnetic measurements as a function of temperature $M(T)$ and field $M(H)$ were performed, along the three high symmetry directions, on the aligned pieces of the single crystal annealed in Ar (3% $H_2$) flow. The susceptibility data were identical to the as grown $Pr_2Zr_2O_7$ crystal, in the temperature range 1.8 to 350 K and in applied fields up to 70 kOe. These results suggest that annealing the as grown crystals in Ar (3% $H_2$) flow for 2 days does not result in any significant changes in the high temperature bulk magnetic behaviour. The calculated values of the effective moment of the Pr ions in our as grown and annealed $Pr_2Zr_2O_7$ crystals are both close to 2.5 $\mu_B$. However, to complete our study, ultra low temperature measurements have to be performed, as those experiments could turn out to be more sensitive to very small variations in chemical composition (<1% of $Pr^{4+}$ reducing to $Pr^{3+}$).

We have also measured the magnetization as a function of temperature on our $Pr_2Zr_2O_7$ crystals grown from starting compositions containing 5% and 10% $Pr^{3+}$ excess, and the Weiss temperatures calculated for these compositions were found to be similar to the previously reported results for $Pr_2Zr_2O_7$ [9], in which a $Pr_6O_{11}$ enriched starting composition was used (exact amount not stated in the paper) (see Table 2).

In comparing the A/B site occupancy ratio and the nominal $Pr^{3+}$ content estimated from the different measurements, we find that the magnetic susceptibility measurements are, however, more sensitive to very small variations of the magnetic ions in the materials than the X-ray diffraction experiments. Fits to the susceptibility data obtained on the crystals show that there is a variation in the Weiss temperatures obtained for the different crystals studied suggesting that there are indeed slight variations in the $Pr^{3+}$ content between samples.

4.       Summary

We have been successful in growing large high quality crystals of the frustrated magnet, $Pr_2Zr_2O_7$ by the floating zone technique, using a growth rate of 15 mm/h and in an oxygen atmosphere, at pressures in the range 1-4 bars. Several crystals have been produced starting with varying levels of excess of $Pr_6O_{11}$ to mitigate the effects of the loss of Pr during the growth process. The high growth rates and gas pressures used during the crystal growths, help to minimise the loss of Pr in our crystals. The lattice parameters increase slightly with the amount of $Pr_6O_{11}$ excess used. X-ray diffraction analysis of the crystals shows the occupancy of the Pr site in the pyrochlore structure to be close to 1.0 for all the crystals grown. The $Pr_2Zr_2O_7$ crystals exhibit a spin-ice type anisotropy. Fits to the susceptibility data show that the Weiss temperatures are negative and that the magnitude of $\Theta_W$ decreases slightly with an increase in the excess of $Pr_6O_{11}$ in the starting material. These changes may be attributed to small variations in the $Pr^{3+}$ content between samples and highlight the sensitivity of this material to chemical composition. Supplementary measurements of the magnetic susceptibility and heat capacity below 1.8 K on $Pr_2Zr_2O_7$ single-crystals may serve to illuminate the spin freezing mechanism of this system. Detailed neutron scattering experiments are now being carried out on our $Pr_2Zr_2O_7$ crystals and structural and magnetic studies on crystal samples of other members of the zirconate family are also in progress.

Acknowledgments

This work was supported by a grant from the EPSRC, UK (EP/I007210/1). The authors thank A. Gukasov, I. Mirebeau, S. Petit and J. Robert for the neutron scattering experiments carried out on our FZ-grown crystals, S. York for the EDAX analysis and T. E. Orton for valuable technical support. Dean S. Keeble thanks the Science City Research Alliance and HEFCE Strategic Development Fund for financial support. Some of the equipment used in this research was obtained through the Science City Advanced Materials Project: Creating and Characterising Next Generation Advanced Materials, with support from Advantage West Midlands (AWM) and part funded by the European Regional Development Fund (ERDF).